\documentclass[11pt]{article}

\usepackage{amsmath,amssymb}
\usepackage{slashed}
\usepackage{xcolor} %Define colors
\usepackage{graphicx}
\usepackage{url}

\newcommand{\be}{\begin{equation}}
\newcommand{\ee}{\end{equation}}
\newcommand{\bea}{\begin{eqnarray}}
\newcommand{\eea}{\end{eqnarray}}
\def\Re{{\cal R \mskip-4mu \lower.1ex \hbox{\it e}\,}}
\def\Im{{\cal I \mskip-5mu \lower.1ex \hbox{\it m}\,}}

\def\tev{\,{\ifmmode\mathrm {TeV}\else TeV\fi}}
\def\gev{\,{\ifmmode\mathrm {GeV}\else GeV\fi}}
\def\mev{\,{\ifmmode\mathrm {MeV}\else MeV\fi}}
\def\to{\rightarrow}

\begin{document}

\begin{center}
\vspace*{15mm}

\vspace{1cm}
{\Large \bf  Higgs production in $e^{-}e^{+}$ collisions as a
  probe of noncommutativity} \\
\vspace{1cm}

{\bf M. Ghasemkhani${}^{1,2}$, R. Goldouzian${}^3$, H. Khanpour${}^{3,4,5}$,
 M. Khatiri Yanehsari${}^{3,6}$, \\
 M. Mohammadi Najafabadi${}^3$}

 \vspace*{.5cm} 
  {\small\sl$^1\,$ Department of Physics, Shahid Beheshti University, G.C., Evin, Tehran 19839, Iran  } \\
   {\small\sl$^2\,$ School of Physics, Institute for Research in Fundamental Sciences (IPM) P.O. Box 19395-5531, Tehran, Iran } \\
   {\small\sl$^3\,$ School of Particles and Accelerators, Institute for Research in Fundamental Sciences (IPM) P.O. Box 19395-5531, Tehran, Iran } \\
 {\small\sl$^4\,$Department of Physics, Farhangian University of Mazandaran, Shariati Branch, Mazandaran, Iran } \\ 
{\small\sl$^5\,$Department of Physics, Faculty of Basic Sciences, Babol University of Technology, P.O.Box 47148-71167, Babol, Iran} \\
 {\small\sl$^6\,$Department of Physics, Ferdowsi University of Mashhad, Mashhad, Iran } 

\vspace*{.2cm}

\end{center}

\vspace*{10mm}
\begin{abstract}
We examine the sensitivity of  the angular distribution of the Higgs boson in
the process of $e^+e^-\to Z H$ and the total cross section in the minimal
noncommutative standard model (mNCSM) framework to set lower limit on the noncommutative
charactristic scale ($\Lambda$). Contrary to the standard model case, 
in this process the Higgs boson tends to be emitted anisotropically in the transverse plane.
Based on this fact,  the profile likelihood ratio
is used to set lower limit on $\Lambda$. The lower limit is presented as a function of the integrated luminosity. 
We show that at the center-of-mass energy of 1.5 TeV and with 500 fb$^{-1}$ of data, the
noncommutative characteristic energy scale $\Lambda$ can be excluded
up to 1.2 TeV.
\end{abstract}

\vspace*{3mm}



%--------------------------------------

\section{Introduction}

The discovery of a  Higgs-like boson by the ATLAS and CMS experiments  \cite{ATLAS}  is considered as
a milestone in illuminating the electroweak symmetry (EW) breaking mechanism.
In case of observing no direct evidence for new physics at the LHC,
one important task would be the precise measurement of the Higgs boson couplings
with the Standard Model (SM) particles. 
In other words, the precise measurement in Higgs sector will be one of the main 
objectives of the future experiments if no new particle is found beside the Higgs boson.
Beside the LHC, it has been shown with detailed realistic simulations that 
the International Linear Collider (ILC) and the Compact Linear
Collider (CLIC) can 
achieve high precision measurements for the Higgs boson properties ~\cite{higgsall,Weiglein:2004hn}. 
It is worth mentioning that the ILC~\cite{ilc} and CLIC
\cite{clic} programs will be to run
at the center-of-mass energies between 200 and 500 GeV to provide the
opportunity for threshold scans like 
$ZH$, $t\bar{t}$, $ZHH$ and $t\bar{t}H$. 
The ultimate goal of the ILC will be increasing the center-of-mass energy to 1
TeV while CLIC aims to reach at the center-of-mass energy of 3 TeV.
Certainly, the LHC will improve the Higgs boson related measurements 
with more data that will be accumulated at the center-of-mass energies of 
13 and 14 TeV. However, it is well known that the precise measurements at the ILC
or CLIC are complementary to the LHC in 
many aspects \cite{higgsall,Weiglein:2004hn,Peskin:2012we}.

As the nature of the space-time may change at Planck scale, a possible generalization
of the ordinary quantum mechanics and quantum field theory to
describe the physics at Planck scale is noncommutativity in space-time.
 Motivations for construction the models on
noncommutative space-time are originating from the string theory, quantum
gravity, and Lorentz breaking \cite{Douglas1,Douglas2,Ardalan2}.

In the simplest way, the noncommutativity can be
described by a set of constant c-number parameters
$\theta_{\mu\nu}$ or equivalently can be charactrized  by an energy scale
$\Lambda$ and dimensionless parameters $C^{\mu\nu}$ as the following:

\begin{eqnarray}
[\hat{x}_{\mu},\hat{x}_{\nu}] = i\theta_{\mu \nu} = \frac{i}{\Lambda^{2}}C_{\mu\nu}
\end{eqnarray}
where $\theta^{\mu \nu}$ is an antisymmetric tensor with the
dimension of $[M]^{-2}$. 
%It is important to notice that a space-time noncommutativity ($\theta_{0i}\neq 0$ ) might lead to violation of unitarity
%and problems to causality \cite{Gomis},\cite{Chaichian}.
%While the unitarity can be satisfied in case of
%$\theta_{0i}\neq 0$ providing that $\theta^{\mu \nu}\theta_{\mu
%\nu} > 0$ \cite{kostelecky}.

A noncommutative version of the ordinary quantum field theory is
obtained only by replacing the ordinary products with the so-called Moyal $\star$
product that is defined as \cite{review,Seiberg99}:
\begin{eqnarray}
(f\star g)(x) &=& \exp\left(\frac{i}{2}\theta^{\mu
\nu}\partial_{\mu}^{y}\partial_{\nu}^{z}\right)f(y)g(z)\bigg\vert_{y=z=x}\\
\nonumber &=&
f(x)g(x)+\frac{i}{2}\theta^{\mu\nu}(\partial_{\mu}f(x))(\partial_{\nu}g(x))+O(\theta^{2}).
\end{eqnarray}

As mentioned, one can construct the noncommutative quantum field theory via Weyl
correspondence in which the ordinary product among the fields is replaced
by the Moyal $\star$ product \cite{Seiberg99}.
To study the noncommutative effects, we concenrtate on the 
minimal version of the noncommutative SM
\cite{Calmet07}. 
By the means of Seiberg-Witten maps, one can expand the matter 
gauge fields in noncommutative space-time 
in terms of the commutative fields as power series of the
noncommutativity parameter $\theta$ \cite{Seiberg99}.
In the approach of Seiberg-Witten maps, the gauge fiels $A_{\mu}$ and matter fields $\psi$ in the noncommutative
space-time can be expanded in terms of the commutative fields as power series of $\theta$:
\begin{eqnarray}
\hat{\psi}(x,\theta)&=&\psi(x)+\theta \psi^{(1)}+...\\
\hat{A}_{\mu}(x,\theta)&=&A_{\mu}(x)+\theta A_{\mu}^{(1)}+...
\end{eqnarray}
In the limit of $\theta \rightarrow 0$ the
noncommutative fields reduce to the fields in the commutative
space-time. One of the interesting advantage of this approach is that it can be
applied to any gauge theory with arbitrary representation of matter field.

The minimal noncommutative SM predicts new interactions among the SM
particles as well as correcting the ordinary
SM vertices. This leads to interesting signals at the collider experiments.
There are already several phenomenological studies on the effects of
noncommutativity on various decay and scattering processes that can be
found in \cite{OHL,Reuter,Madore4,Haghighat,Schupp,Martin1,mojtaba,mojtaba1,Josip1,Josip2,Melic,Namit,Iltan,arfaei,tram1,Hewett01,Hayakawa00,Chai02,Chai03,Das08,Alboteanu07,Ab10,Haghighat06,Buric07,Melic05,ee1,ee2,ee3,ee4,ee5,ee6,ohl2,ohl3,Bilmis,n1,n2}. In most of these studies, lower limits
on the noncommutative charactristic scale have been set.

One interesting effects of noncommutivity is to change the angular distributions of
the final state particles in the scattering processes and in the decay
of unstable particles. It is because of the violation of the angular
momentum conservation in the noncommutative theory.
As an example, in \cite{ee5} it has been shown that the
noncommutativity affects the total cross section and the differential cross sections
significantly in the $e^{-}e^{+}\rightarrow ZH (HH)$ processes.  
Therefore, both the total and differential cross sections
can be used to set lower limit on the noncommutative scale $\Lambda$. 
The same effect is present in $Z\gamma$ production that has 
been studied in \cite{OHL} at the LHC and Tevatron.

As mentioned previously in \cite{ee5} based on theoretical 
calculations, it has been found that the azimuthal distribution of the emitted 
Higgs boson in $e^{-}e^{+}\rightarrow ZH$ process is sensitive to the
noncomutativity. Now, the important task is to perform a more 
realistic study to obtain the possible limits on the noncommutative scale
using this angular distribution
and the total cross section at different center-of-mass energies 
and more importantly at different integrated luminosities of data that will be collected by the
future $e^{-}e^{+}$ experiments. This should be done by employing advanced statistical methods
that are used currently by the large experiments at the LHC and Tevatron.
This will help us to know what would be the outcome of the future experiments 
at different phases of energy and luminosity.

The goal of this short report is to estimate the lower bound on the
noncommutative scale $\Lambda$ at $95\%$ CL in electron-positron
collisions with a Z-boson plus a Higgs boson in the final state.
We set the limit using a test statistics based on the profile likelihood
ratio \cite{PLR} on the angular distribution. We also set 
limit on $\Lambda$ by obtaining the upper limit on the total cross
section of the signal using a Bayesian approach \cite{bayesian}.

This article is organized as follows.
In Section 2, the noncommutative cross section of the
process $e^{-}+e^{+}\rightarrow H+Z$ is shown.
Section 3 is dedicated to use the profile likelihood ratio
to extract the $95\%$ CL lower limit on the noncommutative scale. 
In section 3, we also obtain the expected upper limit on the signal
cross section including $1\sigma$ and $2\sigma$ bands.
We show the limit as a function of the integrated luminosity.

\section{Noncommutative cross section for production of a Higgs boson
  in association with a Z-boson}

\begin{figure}
\centering
 \includegraphics[width=7cm,height=5cm]{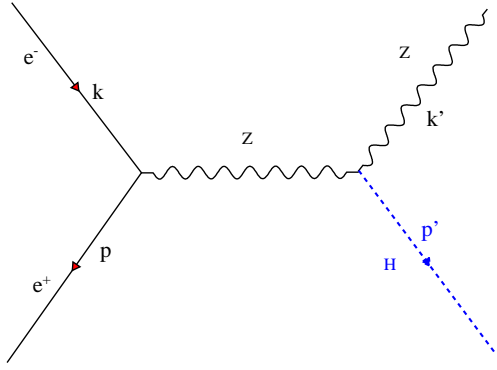}%
 \caption{Feynman diagram for production of a Higgs boson in
 association with a Z-boson in electron-positron collisions.}{\label{feynman}}
 \end{figure}

In this section we show the dependency of the total cross section of 
$e^+e^- \to H Z$ as a function of noncommutative scale ($\Lambda$) as
well as the differential cross section $\frac{d\sigma}{d\phi}$ for
different values of $\Lambda$. This differential distribution has been
proposed in \cite{ee5} to identify the noncommutative effects.  
The Feynman diagram for the $e^+e^- \to H Z$ process is shown in 
Fig.\ref{feynman}. As it can be seen the process proceeds though $s$-channel via
the exchange of a $Z$-boson. It is noteable that the $ZH$ final state can 
be produced via the exchange of Higgs boson that is not considered
because of negiligble electron Yukawa coupling.

The Feynman rule for the vertex $ZZH$ is found to be \cite{ee5}:
\begin{equation}
V_{\mu\nu,ZZH}(p, k,
q)=\frac{im_{Z}^{2}}{v}\{2\cos(\frac{1}{2}p\theta
q)g_{\mu\nu}+\frac{1}{4}((\theta q)_\mu p_{\nu}+(\theta
q)_{\nu}k_{\mu})\times(\frac{\cos(\frac{1}{2}p\theta
q)-1}{p\theta q})\}
\end{equation}
where $m_{Z}$ is the Z-boson mass, $v$ is the vacuum expectation 
value. In the limit of $\theta\rightarrow 0$ the vertex $ZZH$ goes to $\frac{2im_{Z}^{2}}{v}g_{\mu\nu}$ that is
compatible with the SM.
The corresponding matrix element has the
following form:

\begin{equation}
\mathcal{M} \propto \bar{v}(p)\gamma_{\mu}(c_{v}-c_{a}\gamma_5)u(k)\frac{i}{s-m_{Z}^2+i\Gamma_{Z}}V^{\mu\nu}_{ZZH}(k',
p')\epsilon_{\nu}^*(k')e^{\frac{i}{2}p\theta k},
\end{equation}
where  $c_{v}= -\frac{1}{2} + 2 \sin\theta_{W}^{2}, c_{a} = -\frac{1}{2}$ , and $\theta_{W}$ denotes the Weinberg angle.
As it has been shown in Fig.\ref{feynman} $k$, $p$, $k'$ and $p'$ are the four-momenta of electron,
positron, Higgs boson and  Z boson, respectively. 
The center-of-mass energy is denoted by $\sqrt{s} = \sqrt{(k+p)^{2}}=\sqrt{(k'+p')^{2}}$ and  $\Gamma_Z$ is the
width of the $Z$ boson. To calculate the total and differential cross section, 
the equations of motion of the ingoing and outgoing particles are used.
The mass of ingoing particles, $m_{e}$, is ignored in the calculations.

\begin{figure}
\centering
 \includegraphics[width=7cm,height=5cm]{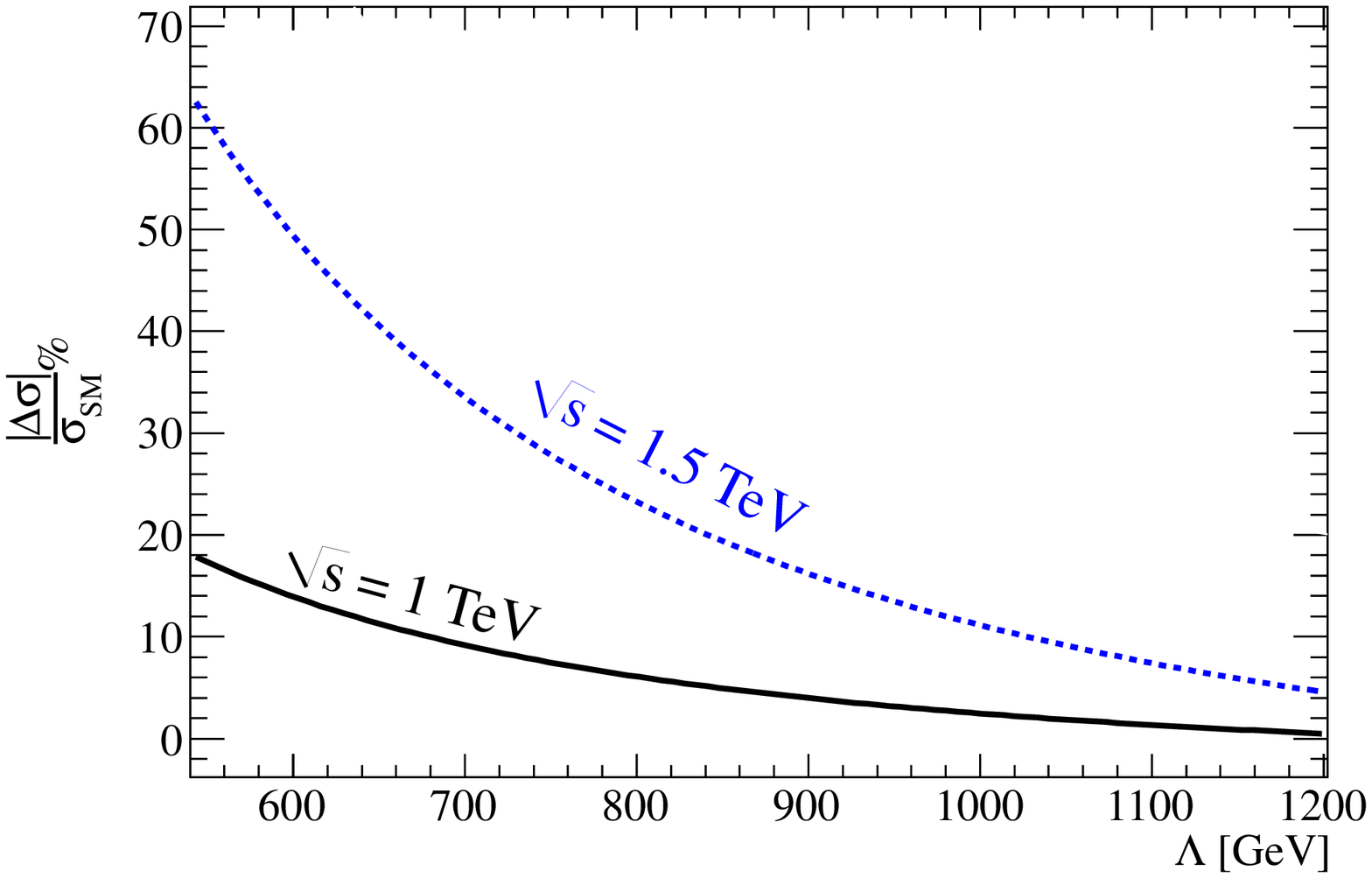}
 \includegraphics[width=7cm,height=5cm]{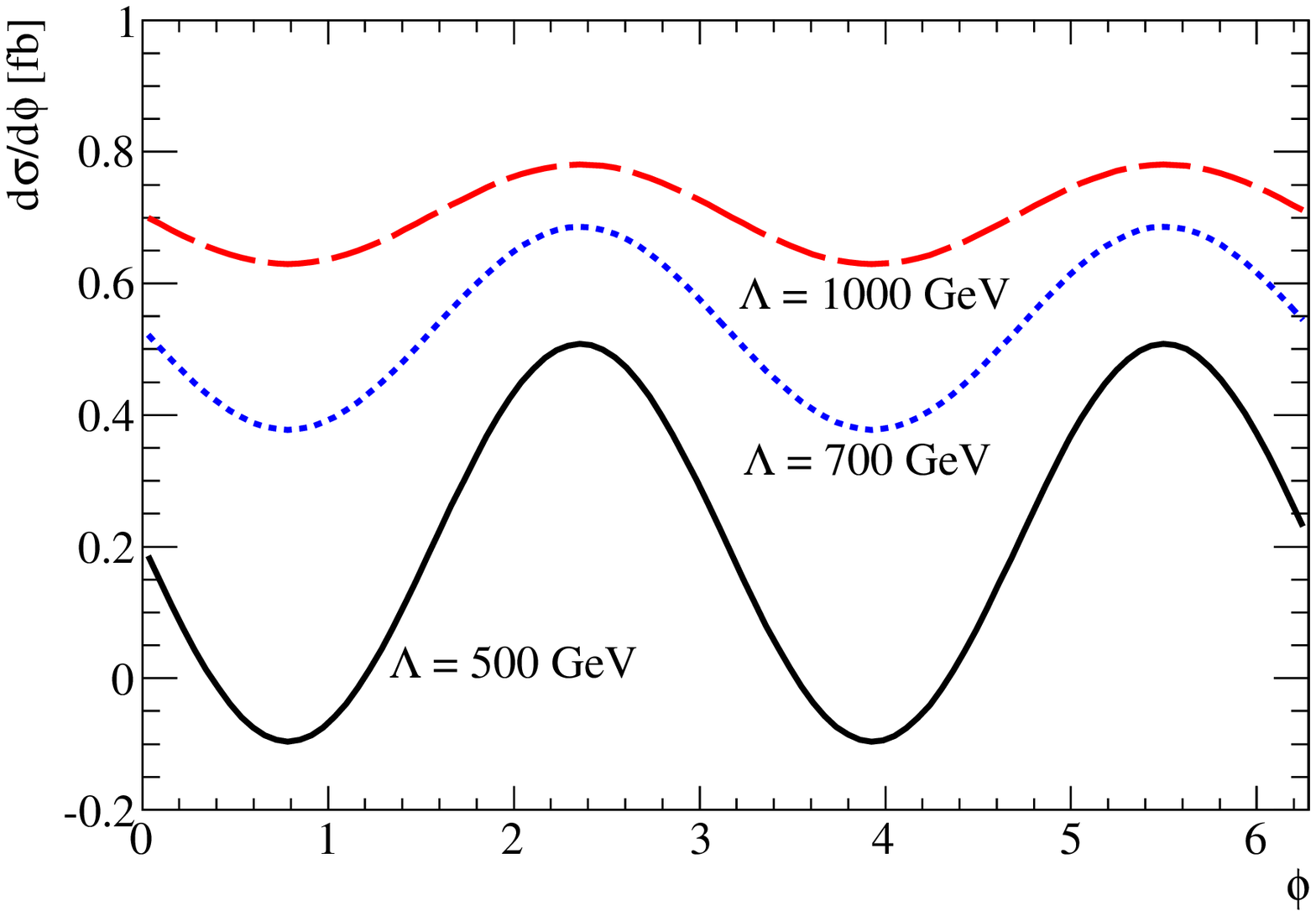}
 \caption{The relative correction from noncommutativity to the cross
 section of $e^+e^-\to Z H$ versus the noncommutative scale
 $\Lambda$ (left) . The differential cross section $\frac{d\sigma}{d\phi}$ for $e^+e^-\to Z H$
 process at various values of $\Lambda$ at $\sqrt{s}=1.5$ TeV (right).}{\label{sigma}}
 \end{figure}

Then the cross section is calculated using the center-of-mass frame for $e^{-}(p) + e^{+}(k) \rightarrow H(p') + Z(k^{'})$ process:
\begin{eqnarray}
p^{\mu} = \frac{\sqrt{s}}{2}(1,0,0, 1) &,& k^{\mu} = \frac{\sqrt{s}}{2}(1,0,0, -1)\\
p'^{\mu } = \frac{\sqrt{s}}{2}(1,\sin\theta\cos\phi, \sin\theta \sin\phi, \cos\theta)&,&
k'^{\mu } = \frac{\sqrt{s}}{2}(1,-\sin\theta\cos\phi, -\sin\theta \sin\phi, -\cos\theta)\nonumber
\end{eqnarray}
where $\theta$ is the polar angle and $\phi$ denotes the azimuthal angle.
After some algebraic manipulations the total and diffenrential cross sections
are obtained. In all calculations in this work the mass of the Higgs boson is 
set to $m_{H}=125$ GeV.  The right plot of Fig.\ref{sigma} shows the
relative correction from noncommutativity to the total cross section
of $e^+e^-\to Z H$ at the center-of-mass energies of 1 and 1.5 TeV
as a function of the noncommutatuve scale $\Lambda$. 
As it can be seen, the noncommutative correction increases
with increasing the center-of-mass energy of the collisions.
%This is expected since the noncommutative corrections 
%in the matrix element are proportional to terms like $(s/\Lambda^{2})^{2}$. 
Because of the significant sensitivity of the total cross section, it can be used to set lower limits on the noncommutative scale.

In addition to the total cross section, on the 
left side of Fig.\ref{sigma} the differential
cross section $d\sigma/d\phi$ at $\sqrt{s}=1.5$ TeV is shown.
From this plot, one can see that contrary to the SM case the distribution of   $d\sigma/d\phi$  
behaves like  $\sin(\phi+\alpha)$.
% where $\alpha$ is a phase that is dependent on the spatial
% orientation of the noncommutativity.
As it can be seen the noncommutativity leads the Higgs boson to be emitted
in an anisotropic way in the transvese plane. This is due to the
violation of the angular momentum conservation in our noncommutatuve model.
An interesting observation is that with increasing the 
noncommutative characteristic scale the amplitude of the
oscillation decreases and goes to zero while the minimum 
and maximum positions do not change.
In the next section, we will set limit on the noncommutative scale 
using the total cross section and the oscillatory behaviour of the emitted Higgs bosons in the transverse plane. 

\section{Test statistics and sensitivity estimates}

\begin{figure}
\centering
 \includegraphics[width=7cm,height=5cm]{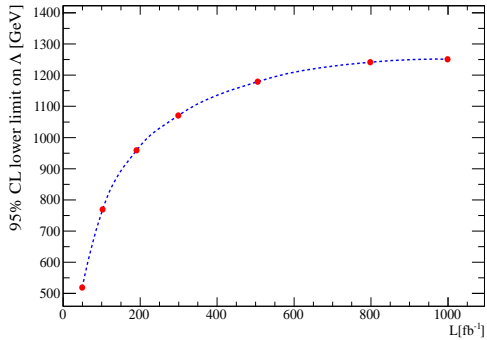}%
 \caption{The $95\%$ C.L lower limits on $\Lambda$ as a function of
 the inegtrated luminosity in electron-positron collider with
 $\sqrt{s}=1.5$ TeV using the azimuthal angular distribution of the Higgs boson without including any systematic effects. }{\label{cls11}}
 \end{figure}

Using the fact that the background, $e^{-}e^{+}\rightarrow ZH$ within 
the SM ($\Lambda \rightarrow \infty$), has a quite
different shape in the $d\sigma/d\phi$ distribution from
the signal (a flat shape versus an oscillating behaviour),
a test statistic can be constructed. Test statistic is a powerful tool 
to separate between signal (noncommutativity) and background and can enhance
the separation power in comparison with other methods. More details can be found in \cite{stat1}.
We use the normalized $\phi$ distribution of the Higgs boson
to define our test statistic as the profile likelihood ratio 
between the two hypotheses of {\it signal+background} and only {\it background}:
\begin{eqnarray}
t=-2ln(\lambda)~\text{with}~\lambda=\frac{L(\mu=1)}{L(\mu=0)}
\end{eqnarray}
where $L$ is defined as:
\begin{eqnarray}
L=\prod_{bin~i}P(n_{i};\mu s_{i}+b_{i})
\end{eqnarray}
where $P$ is the Poisson distribution and $s_{i}$ and $b_{i}$ are the predicted
numbers of signal and background events in bin $i$ of the azimuthal distribution of the Higgs boson, respectively.
We have histogrammed $d\sigma/d\phi$ distribution of the Higgs boson for the signal and background. In each bin of
$\phi$ distribution $s_{i}=s_{i}(\Lambda)=\mathcal{L}\times d\sigma/d\phi_{i}$, in which $\mathcal{L}$ denotes the integrated luminosity.
The quantity $\lambda$ is the profile likelihood ratio which means that for each one of the two values of $\mu=1$ and $\mu=0$,
a fit is performed over the model parameter to find the value which maximizes the likelihood.
The tools for construction of the test statistics have been implemented in the
RooStats framework \cite{roostat} that is a C++ class library based on the RooFit \cite{roofit}
and  ROOT \cite{root} programs. The output is the limit on the model
parameter that is the noncommutative characteristic scale $\Lambda$. 
The results are shown in Fig.\ref{cls11}. It shows the lower limit on the noncommutative
scale as a function of the integrated luminosity. As it can be seen the 
lower limit on $\Lambda$ grows up to around 1.2 TeV when the integrated
lumonisity reaches around 500 fb$^{-1}$. Then with increasing the
integrated luminisity no improvement on the lower limit on $\Lambda$
is observed. This is because of the fact that for $\Lambda \sim 1.2$
TeV and larger values, the oscillating behaviour in $\phi$ distribution
looks like the SM background distribution considering the
uncertainties. It shoud be mentioned here that this estimation is
idealistic as no detector simulation has been performed. In addition,
other backgrounds like $ZZ,W^{+}W^{-}$ have not been considered as
well as all theoretical and instrumental systematic uncertainties.
After considering all the effects one would expect the
limits to be looser. 
To have a rough estimation of the detector and systematic effects
we vary the number of events in each bin of $\phi$ distribution
by $\pm10\%$ to consider these effects as well as the background shape
uncertainty and then recalculate the limit. 
We apply a Gaussian smearing on each bin of the standard model $\phi$ distribution
in order to consider an overall systematic
uncertainties which change the shape of $\phi$ distribution. 
Using $G(m, \sigma)$ a random number that belongs to a Gaussian distribution with
a mean value of $m$ and a standard deviation $\sigma$, the number of
events in each bin of $\phi$ distribution will be smeared as:
\begin{eqnarray}
N_{smeared}(\phi) = \mathcal{L}\times \frac{d\sigma}{d\phi} \times G(1,\Delta)
\end{eqnarray}
where $\mathcal{L}$ is the integrated luminosity and $\Delta$ is set
to $10\%$ as discussed previously.
We found that the lower limit 
on $\Lambda$ decreased to around 1.05 TeV using 500 fb$^{-1}$ of data.   
To obtain a realistic estimation of the sensitivity, 
the backgrounds, detector effects and
selection cuts have to be fully considered. 
The analysis of all backgrounds and simulation of detector effects
is beyond the scope of this short report and must be done
by the experimental collaborations.

\begin{table} 
\begin{center}
\begin{tabular}{|c||c|c|c|}
\hline
        Integrated luminosity        & 100 fb$^{-1}$ & 500 fb$^{-1}$  &    1000 fb$^{-1}$
 \\ \hline 
  limit on $\Lambda$: shape analysis ($d\sigma/d\phi$) & 0.67 TeV  & 1.05 TeV &  1.11 TeV   
 \\  \hline  
   limit on $\Lambda$: total cross section  &   0.62 TeV &  0.94 TeV &   1.10 TeV
\\ \hline 
\end{tabular}
\caption{  The lower limit on the noncommutative scale at two
  interated luminosities of 100, 500 and 1000 fb$^{-1}$ using the shape of
  $\phi$ distribution of the Higgs boson and the total cross section.}
\label{tab}
\end{center}
\end{table}

Another way to set limit on the model parameter is use 
the total cross section of the signal.
By assuming consevative values for the number of
backgrounds and the efficiencies we can set upper limit on the signal 
cross section ($\sigma_{NC}(e^{-}e^{+}\rightarrow ZH)$). Then
the upper limit on the signal cross section can be translated on the
lower bound on the noncommutative scale ($\Lambda$).  
In the absence of a significant excess above the expected background
at any given integrated luminosity of data, one can 
proceed with setting limits on the model parameter $\Lambda$.
To calculate the upper limits on the signal cross section, a counting experiment
is performed. We exploit a standard Bayesian approach
\cite{bayesian} with a flat prior that is chosen for the signal cross
section. More details of the statistical method
can be found in \cite{PLR}. All the calculations are performed with the RooStats \cite{roostat} calculator for
the expected limit.

\begin{figure}
\centering
 \includegraphics[width=7cm,height=5cm]{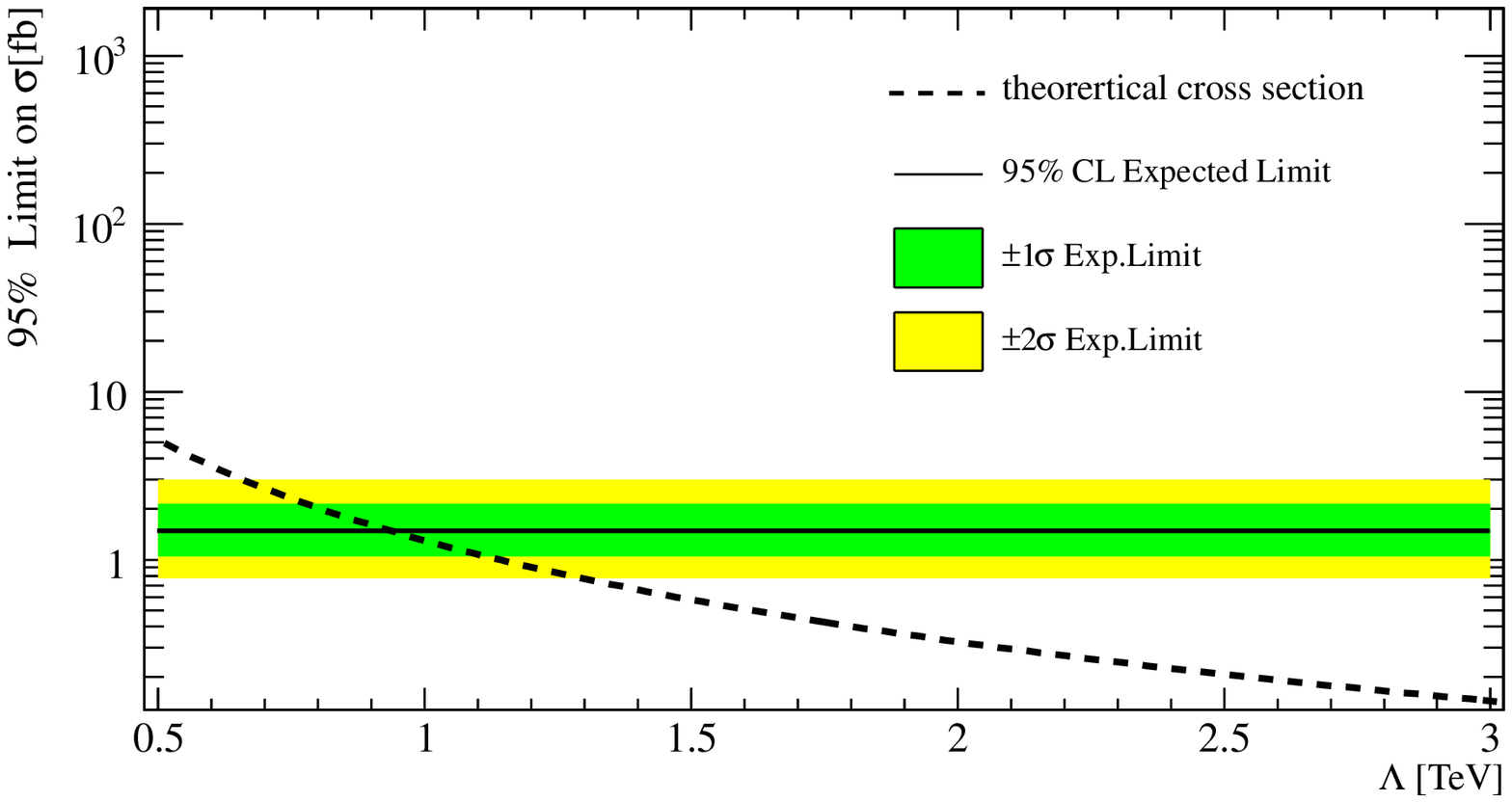}
 \includegraphics[width=7cm,height=5cm]{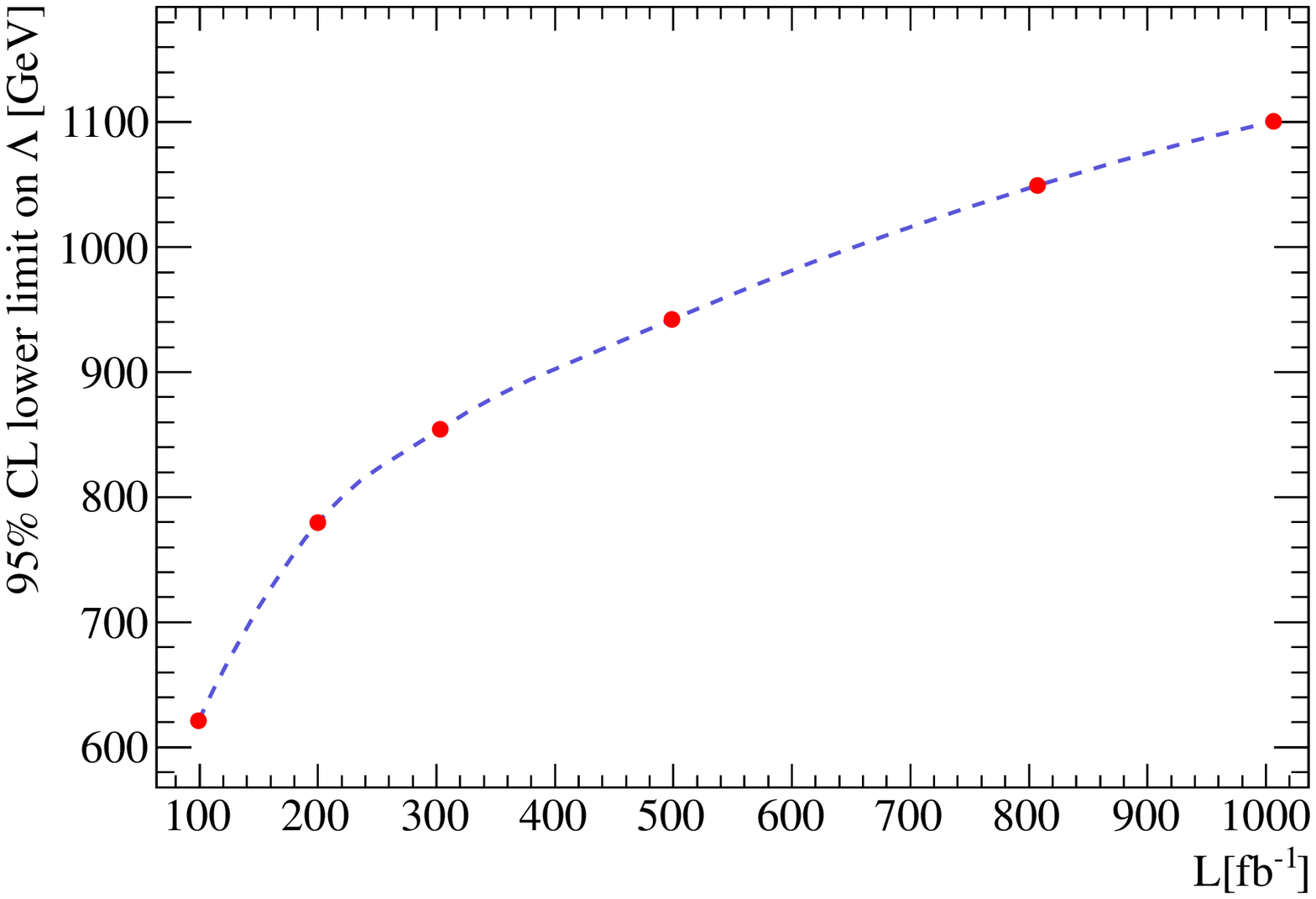}%
 \caption{ The expected $95\%$ CL upper limit on the signal cross
 section using 500 fb$^{-1}$ (left). The expected limit is compared with the theoretical
 prediction. The uncertainty band on the limits at $1\sigma$ and
 $2\sigma$ are shown with shaded bands around the limit. The $95\%$ CL lower limits on $\Lambda$ as a function of
 the inegtrated luminosity in electron-positron collider with
 $\sqrt{s}=1.5$ TeV (right).}{\label{likelihood}}
 \end{figure}

In limit setting process, we choose the conservative numbers of
backgrounds and efficiencies based on
the latest LEP analysis in search for the Higgs boson in $ZH$ cannel \cite{lep}.
We assume the number of survived background events to be twice of the signal
($n_{s}/n_{b}= 0.5$) and an efficiency of signal to be $85\%$ with an
uncertainty of $5\%$. For simplicity, the efficiency is assumed to be fixed
for different values of $\Lambda$.
In the left side of Fig.\ref{likelihood} 
the expected $95\%$ CL upper limit on the signal cross section is
shown. The expected limit is compared with the theoretical
prediction. The uncertainty bands on the limit at $1\sigma$ and
$2\sigma$ are shown with shaded bands around the limit.
The $95\%$ CL lower limits on $\Lambda$ as a function of the inegtrated luminosity
are shown in right side of Fig.\ref{likelihood}. Clearly,
with increasing the amount of data the lower limit on the
noncommutative scale is increased. 
Table \ref{tab} compares the lower bound on $\Lambda$ obtained
from the shape analysis and the upper limit on the signal cross section
for 100, 500 and 1000 fb$^{-1}$ of data at $\sqrt{s}=1.5$ TeV. 

In table \ref{tab1}, we show the lower limit on the noncommutative scale with the
integrated luminosity of 500 fb$^{-1}$ using the shape of
$\phi$ distribution of the Higgs boson and the total cross section for the center-of-mass energies of 0.5,1.0,1.5 TeV.
As it can be seen, the electron-positron collisions with higher center-of-mass energy provide stronger limit on the
noncommutative charactristic scale $\Lambda$.

\begin{table} 
\begin{center}
\begin{tabular}{|c||c|c|c|}
\hline
        $\sqrt{s}$        & 0.5 TeV &  1 TeV  &    1.5 TeV
 \\ \hline 
    limit on $\Lambda$: shape analysis ($d\sigma/d\phi$) & 0.34 TeV  & 0.69 TeV &  1.05 TeV   
 \\  \hline  
   limit on $\Lambda$: total cross section  &   0.31 TeV &  0.63 TeV &   0.94 TeV
\\ \hline 
\end{tabular}
\caption{  The lower limit on the noncommutative scale with the
  integrated luminosity of 500 fb$^{-1}$ using the shape of
  $\phi$ distribution of the Higgs boson and the total cross section for the center-of-mass energies of 0.5,1.0,1.5 TeV.}
\label{tab1}
\end{center}
\end{table}

An interesting feature of noncommutativity that could be considered is to 
study the effect of earth rotation on the cross section of a process. It
leads the cross section to be a time-dependent observable since the detector orientation changes with the earth rotation. 
However, since it is troublesome to have access to the time dependent data for the cross section measurement, the time-averaged
cross section is considered to examine the earth rotation effect.
This average is over the sidereal day which 23 hours, 56 minutes and 4.091 seconds.
For example, in \cite{earth1} the authors have studied the effect of rotation of earth 
on the cross section of $e^{-}e^{+}\rightarrow HH$ process. It has been shown that 
the time-averaged cross section can deviate from the SM prediction around $15\%$.
The earth rotation effect on the  $e^{-}e^{+}\rightarrow HZ$ process also could be
at similar order while precise calculation is needed.
It is interesting to point out here that in \cite{earth2},
the D$\O$ collaboration performed a search for the Lorentz violation 
based on the standard model extension framework (SME) \cite{sme}. Similar to noncomutative SM,
it predicts that the cross sections are dependent on 
sidereal time as the detector orientation changes with the earth rotation. 
The D$\O$ collaboration performed the search 
on the $t\bar{t}$ events based on the SME. Within the uncertainties no time dependent 
effect on the cross section has been observed.

\section{Conclusions}

In this letter we have concentrated on Higgs plus Z-boson production at
a future electron-positron collider to explore the
sensitivity of future accelerator experiments to the noncommutativity.
The noncommutativity destructs the isotropic
azimuthal angular distribution of final state particles.
We used this feature to search for the signal of noncommutative theory
using a test statistic technique. Furthermore, by using a Bayesian
approach with some conservative assumptions for the backgrounds and
efficiencies, conservative estimates obtained on the noncommutative
scale. We find that the Higgs boson angular
distribution shape shows more sensitivity to the model parameter 
than the total cross section. 
It is shown that in this channel,   
the lower limit of 1.1 TeV on $\Lambda$ can be achieved using 500
fb$^{-1}$ of data in electron-positron collisions at the
center-of-mass energy of 1.5 TeV.


\begin{thebibliography}{99}

\bibitem{ATLAS} G.~Aad {\it et al.}  [ATLAS Collaboration],
  %``Observation of a new particle in the search for the Standard Model Higgs boson with the ATLAS detector at the LHC,''
  Phys.\ Lett.\ B {\bf 716}, 1 (2012)
  [arXiv:1207.7214 [hep-ex]]; 
S.~Chatrchyan {\it et al.}  [CMS Collaboration],
  %``Observation of a new boson at a mass of 125 GeV with the CMS experiment at the LHC,''
  Phys.\ Lett.\ B {\bf 716}, 30 (2012)
  [arXiv:1207.7235 [hep-ex]].
  %%CITATION = ARXIV:1207.7235;%%



\bibitem{higgsall}
 S.~Dawson, A.~Gritsan, H.~Logan, J.~Qian, C.~Tully, R.~Van Kooten, A.~Ajaib and A.~Anastassov {\it et al.},
  %``Higgs Working Group Report of the Snowmass 2013 Community Planning Study,''
  arXiv:1310.8361 [hep-ex].
  %%CITATION = ARXIV:1310.8361;%%
  %23 citations counted in INSPIRE as of 07 Mar 2014


\bibitem {Weiglein:2004hn}
G.~Weiglein \textit{et al.} [LHC/LC Study Group Collaboration],
%``Physics interplay of the LHC and the ILC,''
Phys.\ Rept.\ \textbf{426}, 47 (2006) [hep-ph/0410364].
%%CITATION = HEP-PH/0410364;%%

\bibitem{ilc}
H.~Baer, T.~Barklow, K.~Fujii, Y.~Gao, A.~Hoang, S.~Kanemura, 
J.~List and H.~E.~Logan {\it et al.},
  %``The International Linear Collider Technical Design Report - Volume 2: Physics,''
  arXiv:1306.6352 [hep-ph].
  %%CITATION = ARXIV:1306.6352;%

\bibitem{clic} 
  D.~Dannheim, P.~Lebrun, L.~Linssen, D.~Schulte, F.~Simon, S.~Stapnes, N.~Toge and H.~Weerts {\it et al.},
  %``CLIC e+e- Linear Collider Studies,''
  arXiv:1208.1402 [hep-ex];
D.~Dannheim, P.~Lebrun, L.~Linssen, D.~Schulte and S.~Stapnes,
  %``CLIC $e^+ e^-$ Linear Collider Studies - Input to the Snowmass process 2013,''
  arXiv:1305.5766 [physics.acc-ph].




\bibitem {Peskin:2012we}
M.~E.~Peskin,
%``Comparison of LHC and ILC Capabilities for Higgs Boson Coupling Measurements,''
arXiv:1207.2516 [hep-ph].
%%CITATION = ARXIV:1207.2516;%%
%44 citations counted in INSPIRE as of 24 Jun 2013



\bibitem{Douglas1} M. R. Douglas and N. A. Nekrasov, Rev. Mod. Phys. \textbf{73}, 977
(2002).
\bibitem{Douglas2} A. Connes, M. R. Douglas, and A. Schwarz, JHEP \textbf{9802}, 003 (1998).

\bibitem{Ardalan2} F. Ardalan, H. Arfaei and M. M. Sheikh-Jabbari, JHEP \textbf{9902}, 016 (1999).


\bibitem{review} M.R. Douglas, N.A. Nekrasov, Rev. Mod. Phys. \textbf{73}, 977 (2001).
\bibitem{Seiberg99} N. Seiberg, E. Witten , JHEP \textbf{09}, 032 (1999).

\bibitem{Calmet07}X. Calmet, B. Jurco, P. Schupp, J. Wess and M.
Wohlgenannt, Eur. Phys. J. {\bf C23}, 363(2002).

\bibitem{OHL} A. Alboteanu, T. Ohl, R. Ruckl, Phys. Rev. \textbf{D 74}, 096004 (2006).
\bibitem{Reuter} T.~Ohl and J.~Reuter,  Phys.\ Rev.\  D {\bf 70}, 076007 (2004).

\bibitem{Madore4} X. Calmet, B. Jurco, P. Schupp, J. Wess, M. Wohlgenannt, Eur. Phys. J. \textbf{C 23}, 363 (2002).

\bibitem{Haghighat}M. Haghighat, M. M. Ettefaghi, M.
Zeinali, Phys. Rev. \textbf{D 73}, 013007 (2005).
\bibitem{Schupp}P. Schupp, J. Trampetic,
J, Wess and G. Raffelt, Eur. Phys. J. \textbf{C 36}, 405 (2004).
\bibitem{Martin1}C.P. Martin and C.
Tamarit, JHEP \textbf{02}, 066 (2006).
\bibitem{mojtaba}M. Mohammadi Najafabadi, Phys. Rev. \textbf{D 74}, 025021 (2006).
\bibitem{mojtaba1}  S.~Yaser Ayazi, S.~Esmaeili and M.~Mohammadi-Najafabadi,
  %``Single top quark production in $t$-channel at the LHC in Noncommutative Space-Time,''
  Phys.\ Lett.\ B {\bf 712}, 93 (2012)
  [arXiv:1202.2505 [hep-ph]].
  %%CITATION = ARXIV:1202.2505;%%
  %1 citations counted in INSPIRE as of 07 Mar 2014

\bibitem{Josip1} J. Trampetic, [arXiv:0802.2030].
\bibitem{Josip2} M. Buric, D. Latas, V. Radovanovic and J.
Trampetic, Phys. Rev.  \textbf{D 77}, 045031 (2008).
\bibitem{Melic} B. Melic, K.
W. Behr, N.G. Deshpande, G. Duplancic, P. Schupp, J. Trampetic
and J. Wess, Eur. Phys. J. \textbf{C 29}, 441 (2003).
\bibitem{Namit} N. Mahajan, Phys. Rev. \textbf{D 68}, 095001 (2003).
\bibitem{Iltan} E.O. Iltan, Phys. Rev. \textbf{D 66}, 034011 (2002).
\bibitem{Ruckl} A.~Alboteanu, T.~Ohl and R.~Ruckl, Phys.\ Rev.\  D {\bf 76}, 105018 (2007).
\bibitem{arfaei} H. Arfaei, M. H. Yavartanoo, hep-th/0010244.
\bibitem{tram1} 
  J.~Trampetic,
  %``High energy cosmic rays experiments inspired by noncommutative quantum field theory,''
  arXiv:1210.5427 [hep-ph].
  %%CITATION = ARXIV:1210.5427;%%
  %1 citations counted in INSPIRE as of 07 Mar 2014

\bibitem{Hewett01}J.L. Hewett, F. J. Petriello and  T. G.
Rizzo, Phys.Rev. {\bf D64}, 075012(2001); T. G. Rizzo, Int. J.
Mod. Phys. {\bf A18}, 2797(2003).



\bibitem{Hayakawa00}M. Hayakawa, Phys. Lett. {\bf B478},
394(2000).

\bibitem{Chai02}M. Chaichian,  P. Presnajder, M. M. Sheikh-Jabbari and
A. Tureanu, Phys. Lett. {\bf B526}, 132(2002).

\bibitem{Chai03}M. Chaichian, P. Presnajder, M. M. Sheikh-Jabbari and A.
Tureanu, Eur. Phys. J. {\bf C29}, 413(2003).


\bibitem{Das08}P. K. Das, N. G. Deshpande and G. Rajasekaran, Phys. Rev. {\bf
D77}, 035010(2008).

\bibitem{Alboteanu07}A. Alboteanu, T. Ohl and R. Ruckl, Phys. Rev {\bf
D76}, 105018(2007).

\bibitem{Ab10}A. Prakash, A. Mitra and P. K. Das, Phys. Rev. {\bf D82},
055020(2010).

\bibitem{Haghighat06}M. Haghighat, M. M. Ettefaghi and M. Zeinali, Phys.
Rev. {\bf D73}, 013007(2006).

\bibitem{Buric07}M. Buric, D. Latas, V. Radovanovic and J. Trampetic,
Phys. Rev. {\bf D75}, 097701(2007).

\bibitem{Melic05}B. Melic,  K. Passek-Kumericki and J. Trampetic, Phys. Rev
{\bf D72}, 054004(2005).


\bibitem{ee1} 
  W.~Wang, J.~-H.~Huang and Z.~-M.~Sheng,
  %``Bound on Noncommutative Standard Model with Hybrid Gauge Transformation via Lepton Flavor Conserving $Z$ Decay,''
  Phys.\ Rev.\ D {\bf 88}, 025031 (2013)
  [arXiv:1306.1331 [hep-ph]].
  %%CITATION = ARXIV:1306.1331;%%
  %1 citations counted in INSPIRE as of 21 Feb 2014
  \bibitem{ee2} 
  W.~Wang, J.~-H.~Huang and Z.~-M.~Sheng,
  %``TeV Scale Phenomenology of $e^+e^- \to\mu^+ \mu^-$ Scattering in the Noncommutative Standard Model with Hybrid Gauge Transformation,''
  Phys.\ Rev.\ D {\bf 86}, 025003 (2012)
  [arXiv:1205.0666 [hep-ph]].
  %%CITATION = ARXIV:1205.0666;%%
  %2 citations counted in INSPIRE as of 21 Feb 2014
  \bibitem{ee3} 
  N.~G.~Deshpande and S.~K.~Garg,
  %``Anomalous Triple Gauge Boson Couplings in $e^{-}e^{+} \to \gamma \gamma$ for Non Commutative Standard Model,''
  Phys.\ Lett.\ B {\bf 708}, 150 (2012)
  [arXiv:1111.5173 [hep-ph]].
  %%CITATION = ARXIV:1111.5173;%%
  
  \bibitem{ee4} 
  S.~K.~Garg, T.~Shreecharan, P.~K.~Das, N.~G.~Deshpande and G.~Rajasekaran,
  %``TeV Scale Implications of Non Commutative Space time in Laboratory Frame with Polarized Beams,''
  JHEP {\bf 1107}, 024 (2011)
  [arXiv:1105.5203 [hep-ph]].
  
  \bibitem{ee5} 
  W.~Wang, F.~Tian and Z.~-M.~Sheng,
  %``Higgsstrahlung and pair production in $e^{+}e^{-}$ collision in the noncommutative standard model,''
  Phys.\ Rev.\ D {\bf 84}, 045012 (2011)
  [arXiv:1105.0252 [hep-ph]].
  
\bibitem{ee6} 
  S.~Aghababaei, M.~Haghighat and A.~Kheirandish,
  %``Lorentz Violation in the Higgs Sector and Noncommutative Standard Model,''
  Phys.\ Rev.\ D {\bf 87}, 047703 (2013)
  [arXiv:1302.5023 [hep-ph]].

\bibitem{ohl2} 
  A.~Alboteanu, T.~Ohl and R.~Ruckl,
  %``The Noncommutative standard model at the ILC,''
  eConf C {\bf 0705302}, TEV05 (2007)
  [Acta Phys.\ Polon.\ B {\bf 38}, 3647 (2007)]
  [arXiv:0709.2359 [hep-ph]].
  %%CITATION = ARXIV:0709.2359;%%
  %22 citations counted in INSPIRE as of 07 Mar 2014

\bibitem{ohl3} 
  T.~Ohl and C.~Speckner,
  %``The Noncommutative Standard Model and Polarization in Charged Gauge Boson Production at the LHC,''
  Phys.\ Rev.\ D {\bf 82}, 116011 (2010)
  [arXiv:1008.4710 [hep-ph]].
  %%CITATION = ARXIV:1008.4710;%%
  %7 citations counted in INSPIRE as of 07 Mar 2014


\bibitem{Bilmis} 
  S.~Bilmis, M.~Deniz, H.~B.~Li, J.~Li, H.~Y.~Liao, S.~T.~Lin, V.~Singh and H.~T.~Wong {\it et al.},
  %``Constraints on Non-Commutative Physics Scale with Neutrino-Electron Scattering,''
  Phys.\ Rev.\ D {\bf 85}, 073011 (2012)
  [arXiv:1201.3996 [hep-ph]].
  %%CITATION = ARXIV:1201.3996;%%
  %2 citations counted in INSPIRE as of 07 Mar 2014


\bibitem{n1}
 P.~Kumar Das and A.~Prakash,
  %``126 GeV Higgs boson pair production at the linear collider in the noncommutative space-time,''
  Int.\ J.\ Mod.\ Phys.\ A {\bf 28}, 1350004 (2013).
  %%CITATION = IMPAE,A28,1350004;%%


\bibitem{n2}
 P.~K.~Das, A.~Prakash and A.~Mitra,
  %``Neutral Higgs boson pair production at the LC in the Noncommutative Standard Model,''
  Phys.\ Rev.\ D {\bf 83}, 056002 (2011)
  [arXiv:1009.3571 [hep-ph]].
  %%CITATION = ARXIV:1009.3571;%%
  %4 citations counted in INSPIRE as of 23 Jun 2014



\bibitem{PLR}
G. Ranucci, Nucl. Instrum. Meth. A 661 (2012) 77–85;
W. A. Rolke, A. M. Lopez, J. Conrad, Nucl. Instrum. Meth. A 551 (2005) 493–503.

\bibitem{bayesian}
 I. Bertram et al. Fermilab–TM–2104 (2000).


\bibitem{stat1}
G. Cowan, K. Cranmer, E. Gross and O. Vitells, Eur. Phys. J. {\bf C 71}, 1554
(2011), arXiv:1007.1727 [physics.data-an].


\bibitem{roostat} L. Moneta, K. Belasco, K. Cranmer et al. [arXiv:1009.1003 [physics.data-an]].
[https://twiki.cern.ch/twiki/bin/view/RooStats/].

\bibitem{root} R. Brun, F. Rademakers,  Nucl.Instrum. Meth. A389, 81-86 (1997).

\bibitem{roofit} W. Verkerke, D. P. Kirkby, The RooFit toolkit for data modeling, Proceedings for
CHEP03 (2003) [physics/0306116].


\bibitem{lep} 
  M.~M.~Kado and C.~G.~Tully,
  %``The searches for Higgs bosons at LEP,''
  Ann.\ Rev.\ Nucl.\ Part.\ Sci.\  {\bf 52}, 65 (2002).
  %%CITATION = ARNUA,52,65;%%
  %15 citations counted in INSPIRE as of 07 Mar 2014

\bibitem{earth1}
 P.~K.~Das and A.~Prakash,
  %``Effect of earth rotation on pair production of Standard Model Higgs bosons at linear colliders in the noncommutative space-time,''
  Int. J. Mod. Phys. A 28, 1350004 (2013).
  %%CITATION = ARXIV:1207.1246;%%

\bibitem{earth2}
 V.~M.~Abazov {\it et al.}  [D0 Collaboration],
  %``Search for violation of Lorentz invariance in top quark pair production and decay,''
  Phys.\ Rev.\ Lett.\  {\bf 108}, 261603 (2012)
  [arXiv:1203.6106 [hep-ex]].
  %%CITATION = ARXIV:1203.6106;%%
  %17 citations counted in INSPIRE as of 24 Jun 2014

\bibitem{sme}
D. Colladay and V.A. Kostelecky, Phys. Rev. D {\bf 58},
116002 (1998); V.A. Kostelecky, Phys. Rev. D {\bf 69}, 105009 (2004).



\end{thebibliography}
\end{document}